# FORM Matters: Fast Symbolic Computation under UNIX

Michael M. Tung*
Instituto de Matemática Multidisciplinar
Universidad Politécnica de Valencia
P.O. Box 22.012, Valencia, Spain

**Abstract**

We give a brief introduction to FORM, a symbolic programming language for massive batch operations, designed by J.A.M. Vermaseren. In particular, we stress various methods to efficiently use FORM under the UNIX operating system. Several scripts and examples are given, and suggestions on how to use the vim editor as development platform.

**Keywords:** Computer algebra, FORM programming, UNIX

In the past few years, we have experienced rapid progress in the field of computer algebra on desktop PCs. Only two decades ago, many algebraic computations would have been thought impossible to perform on machines other than supercomputer powerhorses. Today, every college student is accustomed to solving his math or physics homework assignments with comfortable computer-algebra environments such as Mathematica, Maple V, Macsyma, and Reduce—just to name the most prominent of such programs.

These modern computer-algebra systems incorporate a vast amount of mathematical knowledge, ranging from extensive integral tables, to obscure properties of special functions and fancy graphical output. However, there are some drawbacks. Although almost all of these mathematical geniuses contain high-precision routines to give numerical output, very few people would seriously employ them for full-scale number crunching in a larger project. For maximum performance, they will rather choose FORTRAN or C/C++ for efficient coding, and link only essential libraries to their programs at runtime. This avoids unwanted overhead and results in executable code which is small and fast.

Similar conditions hold also for a certain class of algebra problems: some algebraic computations rather rely on processing large amounts of symbolic data than making use of encyclopedic mathematical knowledge. For this kind of formula crunching, we require an environment with as little overhead as possible. Here, one has to sacrifice mathematical versatility and sophisticated graphical user interfaces for brute speed and efficiency.

If you have to tackle such problems, consider to use FORM, an algebraic programming language that was specifically designed to manipulate large formulae. Jos A.M. Vermaseren at NIKHEF (the Dutch Institute for Nuclear and High-Energy Physics) created FORM by the end of the 80's to perform large algebraic calculations in his own research, based on a previous project by the Dutch physicist and Nobel prize laureate Martinus J.G. Veltman.

Initially, only the FORM binaries of the first release 1.0 were available to the public for free. For some time later versions 2.x of FORM (introduced in 1991) were commercial and sold by Computer Algebra Netherlands. Now, fortunately and to great benefit of the scientific community, release 3.1 (dated January 30, 2003) is again available for download under the condition that it may not be used for commercial purposes [1].

The latest release 3.1 has many new features and the C code has been substantially rewritten improving speed and efficiency.

In passing from FORM 1 to FORM 2 pattern matching and manipulation was improved, and since then compression is used to store intermediate and final results on disk. With FORM 3 came many additional enhancements, among them most notably dynamical variable administration (with an upper limit of 6000 variables for 32-bit systems!), recursive preprocessor variables and new character string variables [2].

In the following, we guide you through a quick installation of FORM, give you a concise introduction to the basics of its programming language (common to all releases of FORM), and finally show you how to use the vim editor as a powerful workbench for FORM development.

---

*email: mtung@mat.upv.es

## Installation

Installation of FORM is fairly easy. As FORM is written in C, it has been compiled on many UNIX variants (*e.g.* HP, SGI, Solaris, MacOSX, and Alpha architecture).The latest executables for FORM are available for download at NIKHEF's FORM website [1]. The Linux executable is of release 3.1, dated January 2003. It is statically linked for GNU/Linux 2.2.5 and will normally run on older Linux systems. Release versions prior to version 3 are to be found at NIKHEF's ftp site [3, 4]. However, these binaries are dynamically linked to the older libc.so library and will only execute on newer glibc2 systems with backwards compatibility.

After the download, place the binary somewhere into your path, perhaps `/usr/local/bin` and don't forget to set it `chmod 755`. In the following, we will assume that you have named the binary `form3.1`. Test at this stage that you get the following output, when entering `form3.1`:

```
 Correct use is 'form [-options]
↪ inputfile'
```

FORM runs exclusively in batch mode working through a list of commands given in the input file. Unfortunately, there is no `-h` option to present you with more information. The FORM Manual, which is also available in PDF format at NIKHEF's site, gives further details [5].

Fortunately, there are only a few important options that are relevant for setting up things correctly: The option `-s <filename>` specifies where FORM should look for parameter settings. The default filename is `form.set`. If this file is placed in the current directory, where the FORM code to be run is located, it will override all other settings. Thus, only if there is no local `form.set`, it makes sense to use the `-s` option. If no `form.set` file is present at all, FORM will assume built-in defaults, which in most cases will work on standard hardware.

The second option, `-t <pathname>`, defines in which directory FORM should store its temporary files. As FORM can produce enormously huge intermediate files, it is this way possible to redirect temporary output to a partition or device where sufficient physical memory is available. If you have one preferred place to store the temporary file, you don't need to specify this on the command line every time. Instead, use the entry `TempDir` in `form.set`.

The structure of the configuration file is very simple. First comes the keyword of the configuration parameter that is followed by the desired corresponding value, which can be a number, character or string.

The creation of `form.set` is thus straightforward. Here is the minimum configuration we use

```
*
* form.set - global FORM config
*

* setting up temporary directory
TempDir /tmp

* where to look for external code
IncDir  /usr/share/form:/scratch
```

which means that one reserves `/tmp` for temporary storage, and adds the directories `/usr/share/form` and `/scratch` to the search path to look for additional input files outside the execution directory. Commented lines start with an asterisk character. The default comment character could be changed by resetting the configuration keyword `CommentChar`. However, this is not advisable unless needed, since some editors may depend on this default setting.

In older versions of FORM, which assumed PCs with lower memory, it was recommended to set the number of symbols one would work with at a time, and also set the number of statements in one unit of execution. This simple setup should work on most machines. On older PCs, perhaps you will have to tweak some of the parameters that control buffer and heap sizes, most notably `WorkSpace`, `LargeSize`, `SmallSize`, and `MaxTermSize`. In chapter 9 of the user manual, you find all 36 parameters cataloged and explained. If you should have to change any of these parameters to tailor your machine's hardware, consult these guidelines first.

If you have worked out a parameter configuration that you want to make the system-wide default, copy `form.set` to a standard directory as `/etc` or `/usr/local/share` (or perhaps to `~/local/share` for an installation on a particular account without root privileges).

We have written the following wrapper program in the Bourne-Again SHell to assist you when running FORM. The program reads the system-wide parameter defaults (with possible override from local `form.set` files) and provides additional useful features that the FORM binary does not directly implement:

```
#!/bin/bash
#
# form - simple wrapper for FORM
#

FORMBIN=form3.1
FORMSET=/etc/form.set

# not necessary to modify
# anything below this line

if [ "$1" = "" ]; then
  echo "form: You must specify an
  ⇨ input file."
  echo "form <file1> <file2> ..."
  exit
fi

for i in $@; do

FILE=`echo $i | sed s/\.frm$//`
BASE=`basename $1`
DIR=`echo $i | sed s:/$BASE$::`

echo ""

if [ -f $DIR/form.set ] ; then
  $FORMBIN -s $DIR/form.set $i |
  ⇨ tee $FILE.log
 else
  $FORMBIN -s $FORMSET $i |
  ⇨ tee $FILE.log
fi

echo -e "\n FORM output to
⇨ $FILE.log...\n"
```

Make sure that the two environment variables FORMBIN and FORMSET point to the FORM binary and parameter file on your system, respectively. If the binary is placed somewhere in the main path, it won't be necessary to include the full path for FORMBIN. But, you always have to specify the full path for FORMSET.

FORM cannot find a local form.set file when calling an input file from an external directory, and will use the built-in defaults instead. But, calling the wrapper program with

```
form /work/new/inputfile
```

will automatically read a corresponding parameter file form.set in the directory /work/new, if present. The wrapper will also produce a log file in the same directory as the input code, without making use of the -l command-line option of FORM. This way name handling of the log files is much more flexible. If you enter form input.frm (using .frm as the standard extension for FORM) your log output will be input.log. This ensures compatibility with the -l option of FORM 1, which would have produced the more lengthy input.frm.log.

One greater advantage of the wrapper over the naked FORM binary is that it accepts multiple file input. For example, the command line

```
form f1.frm /tmp/f2.frm ~/f3.frm
```

creates log files distributed correctly over the file system, always grouping source and output nicely together. Also for multiple file input, local parameter settings will be used instead of system-wide defaults. This check is done individually for every input file on the command line.

## Programming

FORM sets itself apart through its intentionally restricted set of basic instructions and its terse syntax. Once accustomed to the general syntax principles, you can quite easily extend your working knowledge by looking up advanced or more specialized commands in the FORM Manual and in various tutorials [5, 6, 7]. This section is intended to get you started fast and introduce you to the main features of FORM's programming language. For this purpose, we will treat various mathematical problems and take a closer look on how FORM solves them. As we go on, each problem will present you new aspects and techniques of FORM programming.

### Getting started: Tribonacci numbers

Let's start with a slightly modified classic from computer science that involves recursive programming: Tribonacci numbers [8] are generalized Fibonacci numbers which obey the following recursive relation

$$T_n = T_{n-1} + T_{n-2} + T_{n-3},$$

where the three starting values are $T_1 = 1$, $T_2 = 1$, and $T_3 = 2$. Following this rule, we obtain $T_4$ by taking the sum of the three previous numbers in sequence, namely $T_4 = 1 + 1 + 2 = 4$. To build all higher numbers, we have to proceed in a similar fashion. The FORM program to compute *e.g.* the first 100 Tribonacci numbers will work in the same way. Here is the program that implements this algorithm:

```
*                                      1
* Tribonacci Numbers
*

nwrite statistics;                     5

#define N "100"

Local T1 = 1;
Local T2 = 1;                         10
Local T3 = 2;

#do i = 4, 'N'
  .sort
  drop T'i'-3;                        15
  skip T'i'-2;
  skip T'i'-1;
  Local T'i' = T'i'-1+
               T'i'-2+
               T'i'-3;                 20
  print;
#enddo
.end
```

Note that as for the configuration file the comment character is '*'. Wherever a line begins with an asterisk, the remainder of the line is ignored. This does not conflict with the multiplication sign, as there is always a factor in front of the multiplication '*'. Here, we have given the program a short description at the beginning of the file. Of course, comments can be included in all parts of the coding.

The next observation is that all lines are terminated with a semicolon, except for those with commands that start with the prefixes '.' and '#'. FORM distinguishes between three classes of commands which becomes evident from these punctuation rules. When running a program, the behavior of commands of each class is fundamentally different:

- **Statements and functions** require a terminating semicolon. In our example, these are `nwrite statistics`, `local`, `drop`, `skip`, and `print`. A collection of such statements and functions make up one logical unit in a FORM program. In FORM jargon, such a logical unit is called a *module*.

- **Directives** all start with a period. Hence, in the Tribonacci example, we call `.sort` and `.end` directives. Contrary to statements and functions, directives do not build up the contents of a module, but control modules themselves. All directives denote the end of a module with certain effects. For example, the `.sort` directive terminates the previous module and causes all statements within this unit to be executed. A FORM program is composed of a sequence of modules, where the last module is always closed with the `.end` directive.

- **Preprocessor commands** begin with #, and are exactly what we expect from other programming languages. They are commands to allow for code fragments which are expanded before the actual compilation or execution of a program. In our example, the `#define` command declares that N is to be replaced by 100 at any further occurrence in the remaining program. Note that the argument of `#define` is a string: we have to assign "100" to N and later call the macro with 'N'.

Further important preprocessor commands are `#do` and `#enddo`. These commands help to write your code more compact. Both commands always come in pairs and embrace a code fragment that is to be run several times. In this case, lines 14 to 19 are expanded $N-3$ times, starting with $i=4$ and incrementing up to $i=N$. After this expansion, FORM actually executes the code. Without the do-loop, we would have required $6 \times (N-3)$ lines (not counting additional blank lines) with the same final result!

In FORM all commands are case-insensitive. So, in lines 9–11, we could have written `local` instead of `Local`. Writing commands with initial major letters serves to highlight certain program passages for the human reader. FORM will interpret the program in the same manner, irrespective of the case used for the commands.

On the other hand, variable names and strings are case-sensitive—for the sole purpose to offer the programmer more flexibility. Thus, changing the argument of the `#define` macro in the Tribonacci example from N to n will inevitable result in an execution error.

Use your favorite editor to enter the code of `example1.frm`. Then, run the program by entering `form example1.frm` on the command line. The output will be written to standard output and logged to the file `example1.log`. This is what the log file should contain:

```
FORM by J.Vermaseren,version 3.1
                (Jan 24 2003)
Run at: Thu Jun 10 18:36:44 2004
    *
    * Tribonacci Numbers
    *

    nwrite statistics;
```

```
    #define N "100"

    Local T1 = 1;
    Local T2 = 1;
    Local T3 = 2;

    #do i = 4, 'N'
      .sort
      drop T{'i'-3};
      skip T{'i'-2};
      skip T{'i'-1};
      Local T'i' = T{'i'-1}+
                   T{'i'-2}+
                   T{'i'-3};
      print;
    #enddo

 T4 =
     4;

 T5 =
     7;

 ...

 T99 =
     5332476292809814906472265 8;

  .end

 T100 =
     9807953017858603453650056 4;
```

All programming commands are echoed to output when executed.

On our test hardware (an AMD Athlon XP, 1800 MHz, 1 GB RAM), the execution of the entire program takes on the average about 3 milliseconds. You can check your system's timing by commenting out `nwrite statistics` at the start of the program. Then, in addition to the total timing, FORM will also inform you about the intermediate execution timings, and will provide statistics on the terms generated with their memory usage.

If you have a conventional computer-algebra system at home, try to run the same recursive algorithm and observe how long it takes to obtain this result!

Amazingly, on the same hardware setup, Mathematica 5.0 requires already 28 minutes to calculate $T_{35}$.[1] This is largely due to the memory management of the intermediate results $T_4 \ldots T_{34}$ during the computation. Mathematica and similar algebra systems will store all results in memory, whereas FORM allows to handle memory consumption in the intermediate steps more efficiently.

Let's study the Tribonacci code in more detail and see how FORM achieves such impressive results.

First, we see that the variable `N` is set equal to 100 (line 7), to make later changes more transparent. We could also have supplemented the `#define` command with an appropriate comment for later revisions of the code. In general, it is good programming practice to isolate important variables and document their meaning.

Next, we use in lines 9–10 the `local` statement to define the three local variables $T_1 = 1$, $T_2 = 1$, and $T_3 = 2$. FORM also knows global variables, but they will only be relevant if you have larger output and want to copy results to a special file. In most cases you will probably want to use local variables. The `local` statement serves to declare expressions for FORM to work with. Here, the expressions are purely numerical. In the latter examples, expressions will also be symbolical, involving both numbers and algebraic terms.

Now, let's move on to the core part of the program. As mentioned above, the program block from line 14 to 21 is executed 97 times, each time incrementing $i$ by one. Initially, it is $i = 4$. Then, we sort the results—not much to sort at this point, but it will come in handy later. On line 15, we `drop` the local variable $T_{i-3} \equiv T_1$. With this command $T_1$ is cleared from memory upon the next execution of the `.sort` command. By doing so, we free memory and, in principle, $T_1$ would again be available for reuse.

In the present recursive algorithm, no operations are directly performed on the current $T_{i-2}$ and $T_{i-1}$. We just use these values. By issuing the `skip` command on lines 16 and 17, FORM will keep these variables inactive within the current module. This means that within this range all subsequent commands will have no effect on these variables. In fact, here, we suppress the multiple output of already calculated $T$'s.

Hence, the `print` command of line 21 only acts on the local variable $T_{i=4l}$, which is calculated from its lower three neighbors according to the given recursive relation for each loop $l = 1, 2, \ldots$.

In the last step, after all cycles of the `#do` loop are completed, the `.end` directive gracefully exits the program.

---

[1]After several hours, we lost patience computing $T_{100}$. Mathematica's own extrapolation tool suggests that this calculation will require more than 3 months!

## Pattern matching: Multi-angle trigonometry

Naturally, FORM unleashes its full power when symbolic expressions are involved. Thus, in this example, we will deal with fast symbolic recursion and simple pattern matching.

Since ancient times, mathematical functions have been investigated and charted in huge tables. Usually, the function values for a set of arguments is given and then supplemented by appropriate rules. These rules are designed to extrapolate the results out of documented range.

Let's assume, we would only know the infamous sine and cosine functions for values ranging from 0 to $\pi$ (instead of the full range up to $2\pi$).[2]

We still can recover complete knowledge of the sine with this double-angle formula

$$\sin(2x) = 2\sin(x)\cos(x).$$

Making it more difficult for us, we now suppose knowing only one third, one fourth, ..., one $n$-th of the $2\pi$ range. Nevertheless, we can always expand our knowledge to full scope. Salvation comes from the following recursive relation

$$\sin(nx) = 2\sin[(n-1)x]\cos(x) - sin[(n-2)x].$$

Example 2 implements this algorithm in FORM and computes $\sin(10x)$.

```
*                                          1
* Simplification of
* Multi-Angle Sines
*
                                           5
Symbols x, k, [sin(x)], [cos(x)];
Function sin, cos;

Local expr = sin(10,x);
                                          10
repeat;
  id sin(0,x) = 0;
  id sin(1,x) = sin(x);
  id sin(k?,x) =
         2*sin(k-1,x)*cos(x)              15
                - sin(k-2,x);
endrepeat;

 id sin(x) = [sin(x)];
 id cos(x) = [cos(x)];                    20

print;
.end
```

Execution took just 10 milliseconds on our test machine with the final output

```
Time = 0.01 sec   Generated terms =    407
      expr        Terms in output =      5
                  Bytes used      =    102

expr =
   512*[sin(x)]*[cos(x)]^9 -
   1024*[sin(x)]*[cos(x)]^7 +
   672*[sin(x)]*[cos(x)]^5 -
   160*[sin(x)]*[cos(x)]^3 +
   10*[sin(x)]*[cos(x)];
```

FORM has successfully reduced our symbolic input on line 9 to a sum of powers of simple sines and cosines—there is no more dependence on multiple angles in the arguments.

A look at the program code shows that before working with the local variable `expr`, we have to define the involved symbols and functions (see lines 6–7). Symbols names are composed out of alphanumeric characters, where the first character always has to be a letter. Be careful: FORM is case-sensitive with respect to symbol and function names. Anything between square brackets is not interpreted by FORM. Here, `[sin(x)]` and `[cos(x)]` are symbols as `x` and `k`. Functions are symbols that take arguments.

Lines `11` to `17` are the central part of the program. The `repeat` and `endrepeat` statements embrace command blocks that are to be executed as long as the local variable doesn't change anymore. Our expression is the function `sin(10,x)`. It has two arguments (the first gives the numeric factor $n$ and the second the free variable $x$). The FORM statement `identify` $a = b$ acts on the current local variable(s) and replaces all terms $a$ by $b$. Hence, lines `12` and `13` trivially reduce $\sin(0 \cdot x) = 0$ and $\sin(1 \cdot x) = \sin(x)$. No further decomposition beyond this point is possible.

Lines `14` to `16` demonstrates how pattern matching can be used with the `identify` statement. The first argument `k?` is a placeholder which translates as `k` to the right-hand side of this substitution rule. So, in the initial pass of the `repeat` loop, $\sin(10, x)$ is reduced to combinations of $\sin(9, x)$ and $\sin(8, x)$. In the next run also to $\sin(7, x)$ and $\sin(6, x)$, and so on. The decomposition finally comes to an end with an expression `expr` that only contains immutable terms $\sin(x)$ paired with various factors and $\cos(x)$.

The last two `identify` commands replace the sine and cosine functions by their corresponding symbols.

---

[2]**NB:** Note that there still exist many exotic but in particular cases extremely useful functions that are not yet implemented on computer hardware. In practice, this simple example for the unproblematic sine function can be generalized and applied to these functions.

FORM automatically contracts the powers of symbols, but by default not for functions. Hence, without this trick, the final expression would still contain many uncontracted identical factors.

The program concludes with the usual `print` statement and `.end` directive.

# FORM Development with Graphical Support

FORM programs are run in batch mode. One creates files containing instructions with an editor, let FORM process the file, and investigates the output. During the development of such a program, this edit-and-run cycle is repeated as many times as necessary. Especially for large programs, which often have to run unattendedly, this strategy has proven very efficient.

Our favorite editing tool is the vim editor [9], an improved version of the ubiquitous vi, which adds many powerful features to the original. Apart from its standard editing capabilities, vim can support FORM development by

- using syntax color highlighting;
- executing programs "interactively".

Figure 1 shows a screenshot of the second example program within the graphical mode of vim (enter `gvim example2.frm` on the command line). For automatic syntax highlighting put `syntax on` into the vim configuration file `.vimrc`. Adding the flag

```
let form_enhanced_color=1
```

into `.vimrc` produces a slightly modified coloring that distinguishes better between various FORM command types.

All recent distributions of vim (the current stable release is v6.3) already include FORM syntax highlighting. However, it can also be easily installed on any recent vim installation without doing a full upgrade. In this case, download the syntax file `form.vim` [10] from the vim homepage [9], and follow the instructions in vim's help entry `:help mysyntaxfile`.

Usually, vim detects FORM files by the extension `frm`. Appending to the FORM code the modeline

```
" vim:   ft=frm
```

forces vim to use the `frm` filetype, whatever extension the input file has. The FORM parser will not know about the modeline as it appears after the `end` statement.

While working on a program, it would be quite cumbersome to exit the editor window every time significant modifications are made and you want to run FORM. The following definition in the `.vimrc` shortcuts this tedious edit-and-run cycle and is very helpful for fast prototyping of programs:

```
map ;f :w!<CR>:!form3.1 -l %<CR>
⇨:split `basename % .frm`.log<CR>
⇨:e!<CR><CR>
```

This mapping binds the `;f` key sequence to pass the current file through FORM and display the resulting log file in a split window of the same vim session. To create a menu button in the toolbar of gvim, add, for example, the following line to `.gvimrc`

```
menu 50.10 &FORM.run\ form3.1  ;f
```

The numbers `50.10` indicate the location in the toolbar and will depend on your local vim installation. Take a look at the global settings in `menu.vim` to find suitable values that fit your system. For example, you could easily add another command with the entry `menu 50.20 ...`.

Figure 2 gives a screenshot of gvim running FORM on `example2.frm`. The upper part of the screen displays the corresponding log output in a split window.

# Outlook

This has been a brief introduction to FORM and with emphasis on how to use it effectively under UNIX. FORM offers many more features to perform advanced and large algebraic computations. Most notably, with time and experience you can build up your own libraries of procedures that contain specialized routines which solve regularly occurring mathematical tasks.

# References and Resources

[1] NIKHEF FORM Website, The National Institute for Nuclear Physics and High Energy Physics, Amsterdam. The FORM license and latest release can be found at website `http://www.nikhef.nl/~form/license.html`.

[2] J.A.M. Vermaseren, *New features of FORM*, arXiv:math-ph/0010025 v2, Amsterdam (2000).

[3] NIKHEF Ftp Repository at `ftp://ftp.nikhef.nl/pub/form/`. Older FORM v1.x binaries for most common UNIX flavors (and of course Linux) can be downloaded from this ftp site. The original FORM manual and other shorter introductions are also located at this site.

[4] J.A.M. Vermaseren, *FORM*, 252p., Amsterdam (1989). This is the original reference to FORM v1.x.

[5] J.A.M. Vermaseren, *Symbolic Manipulation with FORM, Tutorial and Reference Manual*, 113p., Amsterdam (2002). The new handbook that deals with FORM 3. It is a revised and enlarged version of the previous reference manual for FORM 2.

[6] A. Heck, *FORM for Pedestrians*, CAN Expertise Center and University of Petrópolis, Petrópolis (1993); *ibid.*, with J.A.M. Vermaseren, 149p., Amsterdam (2000). An introduction with many good examples and exercises. It was updated and extended to FORM 3.

[7] G.J. van Oldenborgh, *An Introduction to FORM*, 26p., Leiden (1995). A concise introduction with some advanced math and physics applications.

[8] S. Plouffe, *Plouffe's Tables of Constants*, University of Quebec, Montreal (1999). A short note on how to calculate Tribonacci numbers without recursive algorithms is located at `http://pi.lacim.uqam.ca/piDATA/tribo.txt`.

[9] The Vim Homepage, located at `http://www.vim.org`. Binaries and source code are available for download.

[10] M.M. Tung, *Vim coloring scheme for syntax elements in FORM*, Mainz (2001); located at `http://www.vim.org/htmldoc/syntax.html#form.vim` and `ftp://ftp.home.vim.org/pub/vim/runtime/syntax`.

```
* Simplification of Multi-Angle Sines
*
Symbols x, k, [sin(x)], [cos(x)];
Function sin, cos;

Local expr = sin(10,x);

repeat;
  identify sin(0,x) = 0;
  identify sin(1,x) = sin(x);
  identify sin(k?,x) = 2*sin(k-1,x)*cos(x)
                     - sin(k-2,x);
endrepeat;

id sin(x) = [sin(x)];
id cos(x) = [cos(x)];

print;
.end
```

Figure 1: Multi-angle sine example with vim syntax highlighting.

Figure 2: Running and logging of the multi-sign example with vim